\documentclass[twocolumn]{aastex63}

\shorttitle{Point sources at $z\sim8$}

\usepackage{color}		
\definecolor{midgray}{gray}{0.4}	
\definecolor{orange}{rgb}{1,0.5,0} 
\definecolor{blue}{rgb}{0,0,0.6}  
\usepackage{apjfonts}

\definecolor{ao}{rgb}{0.0, 0.0, 1.0}

\usepackage{scrextend}
\deffootnote[0.5em]{0em}{1em}{\textsuperscript{\thefootnotemark}\,}

\usepackage{etoolbox}
\makeatletter
\patchcmd{\NAT@citex}
  {\@citea\NAT@hyper@{\NAT@nmfmt{\NAT@nm}\NAT@date}}
  {\@citea\NAT@nmfmt{\NAT@nm}\NAT@hyper@{\NAT@date}}
  {}
  {}
\patchcmd{\NAT@citex}
  {\@citea\NAT@hyper@{%
     \NAT@nmfmt{\NAT@nm}%
     \hyper@natlinkbreak{\NAT@aysep\NAT@spacechar}{\@citeb\@extra@b@citeb}%
     \NAT@date}}
  {\@citea\NAT@nmfmt{\NAT@nm}%
   \NAT@aysep\NAT@spacechar%
   \NAT@hyper@{\NAT@date}}
  {}
  {}
\patchcmd{\NAT@citex}
  {\@citea\NAT@hyper@{%
     \NAT@nmfmt{\NAT@nm}%
     \hyper@natlinkbreak{\NAT@spacechar\NAT@@open\if*#1*\else#1\NAT@spacechar\fi}%
       {\@citeb\@extra@b@citeb}%
     \NAT@date}}
  {\@citea\NAT@nmfmt{\NAT@nm}%
   \NAT@spacechar\NAT@@open\if*#1*\else#1\NAT@spacechar\fi%
   \NAT@hyper@{\NAT@date}}
  {}
  {}
\makeatother

\makeatletter
\def\blfootnote{\xdef\@thefnmark{}\@footnotetext}
\makeatother

\newcommand{\myemail}{tmorishita@stsci.edu}
\newcommand{\simgt}{\,\rlap{\lower 3.5 pt \hbox{$\mathchar \sim$}} \raise
1pt \hbox {$>$}\,}
\newcommand{\simlt}{\,\rlap{\lower 3.5 pt \hbox{$\mathchar \sim$}} \raise
1pt \hbox {$<$}\,}

\newcommand{\logZ}{\log Z_*/Z_\odot}
\newcommand{\spbg}{{SuperBoRG}}
\newcommand{\ly}{${\rm Ly\alpha}$}

\newcommand{\hb}{${\rm H\beta}$}

\newcommand{\yamato}{sBoRG-0853+0310-258}

\newcommand{\ew}{3000}
\newcommand{\fesc}{2}

\newcommand{\kms}{{\rm km~s^{-1}}}
\newcommand{\oii}{[\textrm{O}~\textsc{ii}]}
\newcommand{\oiii}{[\textrm{O}~\textsc{iii}]}

\newcommand{\ciii}{\textrm{C}~\textsc{iii}]}

\def\numfld{295} 
\def\Afld{$0.4\,\mathrm{deg}^2$} 
\def\num{3} 
\def\numall{30} 

\newcommand{\hst}{{HST}}

\newcommand{\spit}{{Spitzer}}

\newcommand{\Z}{{\rm [Z/H]}}

\newcommand{\galfit}{{\ttfamily GALFIT}}
\newcommand{\sext}{{SExtractor}}

\newcommand{\eazy}{{\ttfamily EAzY}}

\newcommand{\gsf}{{\ttfamily gsf}}
\newcommand{\fsps}{{\ttfamily fsps}}

\newcommand{\affilA}{Space Telescope Science Institute, 3700 San Martin Drive, Baltimore, MD 21218, USA; \href{mailto:\myemail}{\myemail}}
\newcommand{\affilB}{School of Physics, Tin Alley, University of Melbourne, VIC 3010, Australia}
\newcommand{\affilC}{ARC Centre of Excellence for All Sky Astrophysics in 3 Dimensions (ASTRO 3D), Australia}
\newcommand{\affilD}{Department of Physics and Astronomy, UCLA, 430 Portola Plaza, Los Angeles, CA 90095-1547, USA}
\newcommand{\affilE}{Center for Astrophysics | Harvard \& Smithsonian, 60 Garden Street, Cambridge, MA, 02138, USA}
\newcommand{\affilF}{Tomonaga Center for the History of the Universe (TCHoU), Faculty of Pure and Applied Sciences, University of Tsukuba, Tsukuba, Ibaraki 305-8571, Japan}
\newcommand{\affilG}{Department of Physics \& Astronomy, Johns Hopkins University, Bloomberg Center, 3400 North Charles Street, Baltimore, MD 21218, USA}

\begin{document}
\title{ \Large
SuperBoRG: Exploration of point sources at $z\sim8$ in HST parallel fields$^{\dagger}$
}

\blfootnote{$\dagger$ Based on observations made with the NASA/ESA Hubble Space Telescope, \\
obtained from the data archive at the Space Telescope Science Institute. STScI \\
is operated by the Association of Universities for Research in Astronomy, Inc. \\
under NASA contract NAS 5-26555 (doi: 10.17909/t9-m7tx-qb86).}

\author{T.~Morishita}
\affiliation{\rm \affilA}

\author{M.~Stiavelli}
\affiliation{\rm \affilA}

\author{M.~Trenti}
\affiliation{\rm \affilB}
\affiliation{\rm \affilC}

\author{T.~Treu}
\affiliation{\rm \affilD}

\author{G.~W.~Roberts-Borsani}
\affiliation{\rm \affilD}

\author{C.~A.~Mason}
\affiliation{\rm \affilE}
\affiliation{\rm Hubble Fellow}

\author{T.~Hashimoto }
\affiliation{\rm \affilF}

\author{L.~Bradley}
\affiliation{\rm \affilA}

\author{D.~Coe}
\affiliation{\rm \affilA}

\author{Y.~Ishikawa}
\affiliation{\rm \affilG}

\received{2020 August 17}
\revised{2020 September 9}
\accepted{2020 September 14}


\begin{abstract}
\noindent
To extend the search for quasars in the epoch of reionization beyond the tip of the luminosity function, we explore point-source candidates at redshift $z\sim8$ in \spbg, a compilation of $\sim$\,\Afld archival medium-deep ($m_{\rm F160W}\sim 26.5$\,ABmag, 5$\sigma$) parallel infrared (IR) images taken with the Hubble Space Telescope (HST). Initial candidates are selected by using the Lyman-break technique. We then carefully analyze source morphology, and robustly identify \num\,point sources at $z\sim8$. Photometric redshift analysis reveals that they are preferentially fit by extragalactic templates, and we conclude that they are unlikely to be low-$z$ interlopers, including brown dwarfs. A clear IRAC ch2 flux excess is seen in one of the point sources, which is expected if the source has strong \hb+\oiii\ emission with a rest-frame equivalent width of $\sim\ew$\,\AA. Deep spectroscopic data taken with Keck/MOSFIRE, however, do not reveal \ly\ emission from the object. In combination with the estimated \hb+\oiii\ equivalent width, we place an upper limit on its \ly\ escape fraction of $f_{\rm esc, Ly\alpha}\simlt \fesc \%$. We estimate the number density of these point sources as $\sim1\times10^{-6}$\,Mpc$^{-3}$\,mag$^{-1}$ at $M_{\rm UV}\sim-23$\,mag. The final interpretation of our results remains inconclusive: extrapolation from low-$z$ studies of {\it faint} quasars suggests that $\simgt100\times$ survey volume may be required to find one of this luminosity. The James Webb Space Telescope will be able to conclusively determine the nature of our luminous point-source candidates, while the Roman Space Telescope will probe $\sim 200$ times the area of the sky with the same observing time considered in this HST study.
\end{abstract}



%
\section{Introduction}\label{sec:intro}
The formation of the first generation of quasars and galaxies is one of the top priority areas in current astronomical research. In addition to their formation mechanism itself, connected to the growth of cosmic structure, understanding their number density evolution and properties during the first billion years is critical to answer the question --- which sources (re-)ionized the universe, and how did they do it \citep{robertson15,bouwens15,madau15}?

Our understanding of early star, galaxy and black hole formation has been significantly improved in the past decade. The installation of Wide Field Camera 3 (WFC3) on \hst\ has enabled us to explore beyond the previous redshift limit of $z\sim 6.5$, related to the observer-frame Lyman break moving outside the sensitivity of silicon-based imaging sensors and into a region of elevated atmospheric foreground. We now have hundreds of galaxy candidates at $z\simgt7$ from multiple legacy surveys of \hst, such as the Cosmic Assembly Near-infrared Deep Extragalactic Legacy Survey \citep[CANDELS;][]{koekemoer11,grogin11}, the Hubble Ultra Deep Field 2012 \citep{ellis13}, the eXtreme Deep Field \citep{illingworth13}, and the Hubble Frontier Fields \citep{lotz17}. Follow-up spectroscopic campaigns, however, confirmed only a portion ($\sim20$) of those candidates at $z>7$, exhibiting the challenging aspect of identifying objects via the \ly\ line at this early epoch \citep{treu13,schenker14,song16}, due to increasing fraction of neutral hydrogen \citep{konno14,mason18,banados18,davies18,hoag19}. 

While low escape fraction of \ly\ photons is theoretically expected and observationally seen in typical galaxies ($L<L^*$) at such an early epoch owing to neutral gas in proximity of these sources, the situation may be different for luminous sources. Theoretically, luminous sources are able to create a large ionizing bubble, where a higher fraction of \ly\ photons can escape \citep{cen00}. A few of such examples with very high escape fraction are indeed seen at $z>6$ \citep{matthee18,tilvi20}. Therefore, luminous objects are ideal targets where it should be more likely to detect \ly\ emission, thus enabling investigation of at least some of their properties \citep{mason18b}. 

However, such luminous sources are rare, and significantly affected by cosmic variance \citep[][]{trenti08}. As an example, \citet{roberts-borsani16} identified three $z\sim8$ galaxies in one of five fields in CANDELS, EGS, whereas only one from the other four fields. \citet[][]{tilvi20} recently revealed an over density of three galaxies at $z=7.7$, again in the EGS field. This demonstrates that a survey over hundreds of independent sightlines is highly complementary to large-area legacy surveys that observe with a mosaic strategy along a small number of sightlines. The Brightest of Reionizing Galaxies survey \citep[BoRG;][]{trenti11,bradley12}, is one of such surveys, which comprises of multi-band imaging along random-pointing fields selected from high galactic latitude pure-parallel opportunities. Indeed, previous campaigns of BoRG have successfully collected $z\simgt8$ galaxy candidates at the bright-end, $M_{\rm UV}\sim-21$ to $-24$\,mag, offering robust determination of the luminosity function at $L>L_*$ \citep{schmidt14,calvi16,morishita18b,livermore18}.

So far, the BoRG survey data have been used to select primarily clearly resolved galaxy candidates, and high-$z$ candidates appearing as point sources were excluded. This is also the case for many other studies, as those point sources are considered to be most likely foreground low-mass stars (i.e. brown dwarfs), given these local sources also have a strong spectral break at $\simlt1\mu{\rm m}$. However, while this is true for $z\simlt7$ galaxy candidates \citep[e.g.,][]{pirzkal05}, the Lyman break shifts towards longer wavelength at yet {\it higher redshift}, and thus quasar/compact galaxy colors become separable from typical brown dwarfs. Thus revisiting existing \hst\ data sets to identify high-$z$ candidates with point-source morphology, especially at the bright end where extremely luminous galaxies and low-luminosity quasars co-locate, provides an ideal opportunity to investigate the nature of extreme starburst at high redshift, {and potentially the number density of low-luminosity quasars too.}

Theoretical studies expect that quasar luminosity functions (LFs) evolve with redshift to different functional form depending on the mode of black hole evolution \citep[e.g.,][]{volonteri10,ren20}. Currently, wide-field surveys have characterized their LFs up to $z\sim7$ primarily from the ground, and only at the very bright end, $M_{\rm UV}<-24$\,mag \citep[][]{jiang09,willott10,fiore12,giallongo15,kashikawa15,matsuoka18b}. However, its exploration at higher redshift, where LFs from different models (e.g., Schechter versus double power law) deviate more significantly, is hampered by increasing atmospheric foreground in near-infrared imaging from the ground. Therefore, space-based observatories are critically needed, and currently the only choice for this purpose is \hst.

In this paper, we aim to identify such luminous high-$z$ point-source candidates and set limits on their number density. To do so we compiled multi-band imaging data from the \hst\ archive for all suitable {parallel} observations in the past decade, including {\it pure-parallel} and {\it coordinated-parallels} surveys. From these surveys, we have collected multi-band imaging data from \numfld\,sightlines, reaching $\sim$\,\Afld. While the overview of the project, \spbg, and its data analysis details are provided in \citet{morishita20b}, in this study we primarily focus on $z\sim 8$ candidates, selected by means of Lyman break technique and consistent with point-source morphology.

The paper is structured as follows. In Section~\ref{sec:data}, we describe the data set and select point-source candidates at $z\sim8$. We investigate their photometric and spectroscopic nature in Section~\ref{sec:nature}. In Section~\ref{sec:lum}, we take a deeper look at one of the point sources with spectroscopic non-detection of \ly, and then estimate number densities of point sources. Throughout, we quote magnitudes in the AB system, and we assume $\Omega_m=0.3$, $\Omega_\Lambda=0.7$, $H_0=70\,\kms\, {\rm Mpc}^{-1}$.

\begin{figure*}
\centering
	\includegraphics[width=0.95\textwidth]{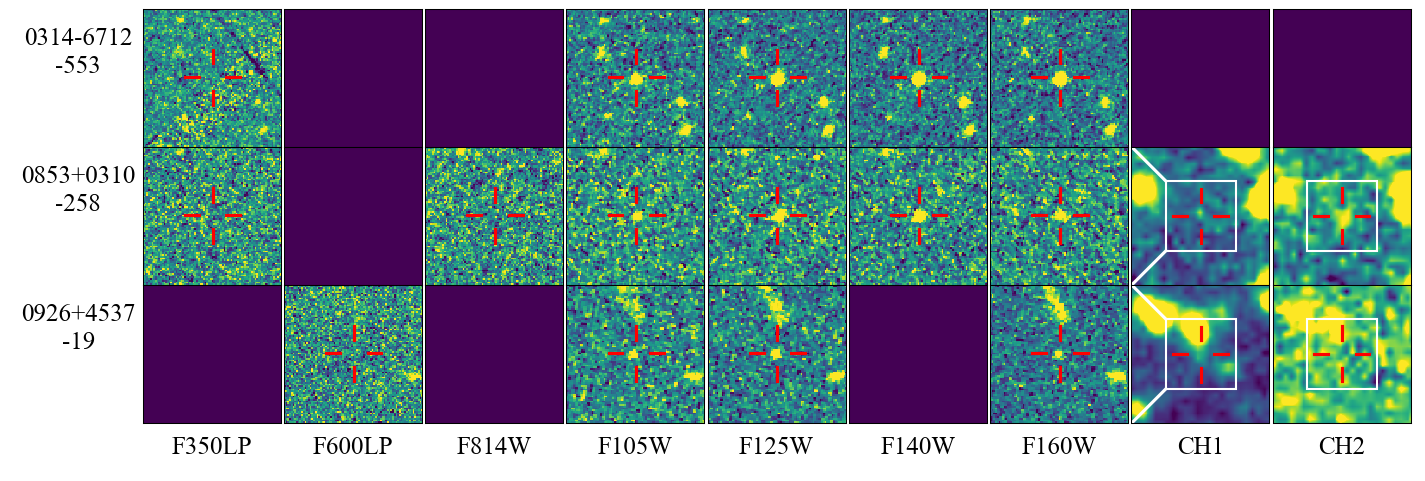}
	\caption{
	Three point sources selected from the photometric $z\sim8$ candidates. Postage stamps are $6\arcsec\!.5\times6\arcsec\!.5$ in size for \hst\ and $12\arcsec\!.8\times12\arcsec\!.8$ for \spit\ bands. The size of \hst\ stamps is shown in \spit\ images (white rectangles). Columns where filters are not available are left blank.
	}
\label{fig:stamp_z8Y105}
\end{figure*}

\section{Data and Sample Selection}\label{sec:data}

\subsection{Data: HST pure-parallel observations}\label{ssec:data}
We use a photometric catalog constructed in the \spbg\ project \citep{morishita20b}. Briefly, \spbg\ compiles several \hst\ parallel surveys, including {\it pure-parallel} programs such as BoRG \citep[][G.~W.~Roberts-Borsani, in prep., N.~Leethochawalit, in prep., from \hst\ cycles 17,19,22,25]{trenti12,bradley12,schmidt14,calvi16,morishita18b,livermore18}, Hippies \citep[][]{yan11}, and coordinated parallels of CLASH \citep{postman12} and RELICS \citep{salmon20}, spanning over \numfld\,independent sightlines with WFC3 multi-band images. These imaging data are medium-deep (typically $\sim2\,k$\,sec to $5\,k$\,sec in each filter), consist of optical to NIR images, and are therefore optimal for searching for high-$z$ luminous galaxy and quasar candidates.

We reduce all \hst\ data with the new version of our custom pipeline, that maximizes science image quality for non-dithered images. The major update since \citet{calvi16} and \citet{morishita18b} is an improved background subtraction, where the new version models local background fluctuations across the detectors. This improves the limiting magnitude by $\sim0.1$\,mag in all filters. In building source catalogs through SourceExtractor \citep{bertin96}, we also switch to aperture photometry from previously used isophotal photometry. This improves the signal-to-noise ratio of blue bands, which are critical to characterize the Lyman break, and significantly decreases the fraction of low-$z$ interloper misidentified as dropout candidates.

In addition, for some \spbg\ fields, \spit/IRAC data are available from a variety of observing programs. Those were primarily taken as part of follow-up campaigns of high-$z$ candidates identified previously, but a few fields also have moderately deep images taken as part of other independent investigations that happen to (partially) overlap with the \hst\ fields considered here. For every \spbg\ field we check availability of data in the IRAC archive and, if available, we download images in the level 1 (bcd) format and combine them with the IRAC pipeline to the level 2 (pbcd) format.

\begin{figure}
\centering
	\includegraphics[width=0.45\textwidth]{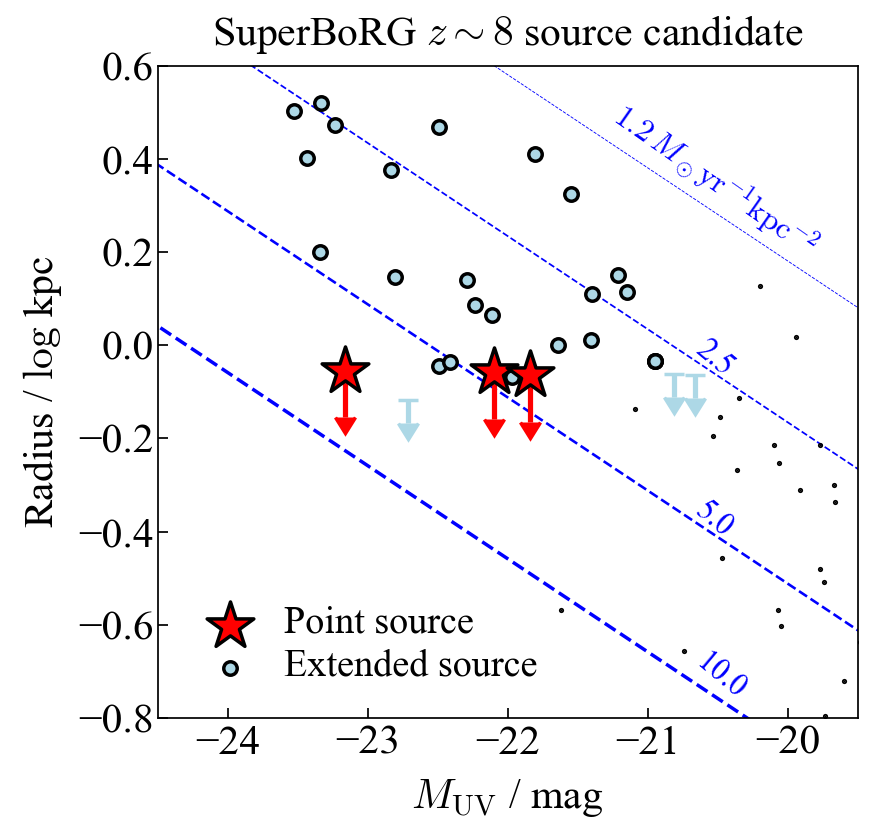}
	\caption{
	Sources selected as $z\sim8$ candidates (z8\_Y105) from \spbg\ shown on the UV absolute magnitude-size diagram. All of the \num\ sources selected with our point-source criteria (red star symbols; Sec.\ref{ssec:sele}) are not resolved and have upper limits in size (although note that size itself is not used for the point-source selection, nor corrected for PSF effects). Most of other z8\_Y105 candidates (blue circles) are resolved, with only three unresolved (blue arrows; see Sec.\ref{ssec:sele}), and in line with the relation from fainter, lens-magnified, galaxy candidates at $z\sim8$ \citep[black dots;][]{kawamata18}.
	}
\label{fig:Mr}
\end{figure}

\begin{figure*}
\centering
	\includegraphics[width=0.33\textwidth]{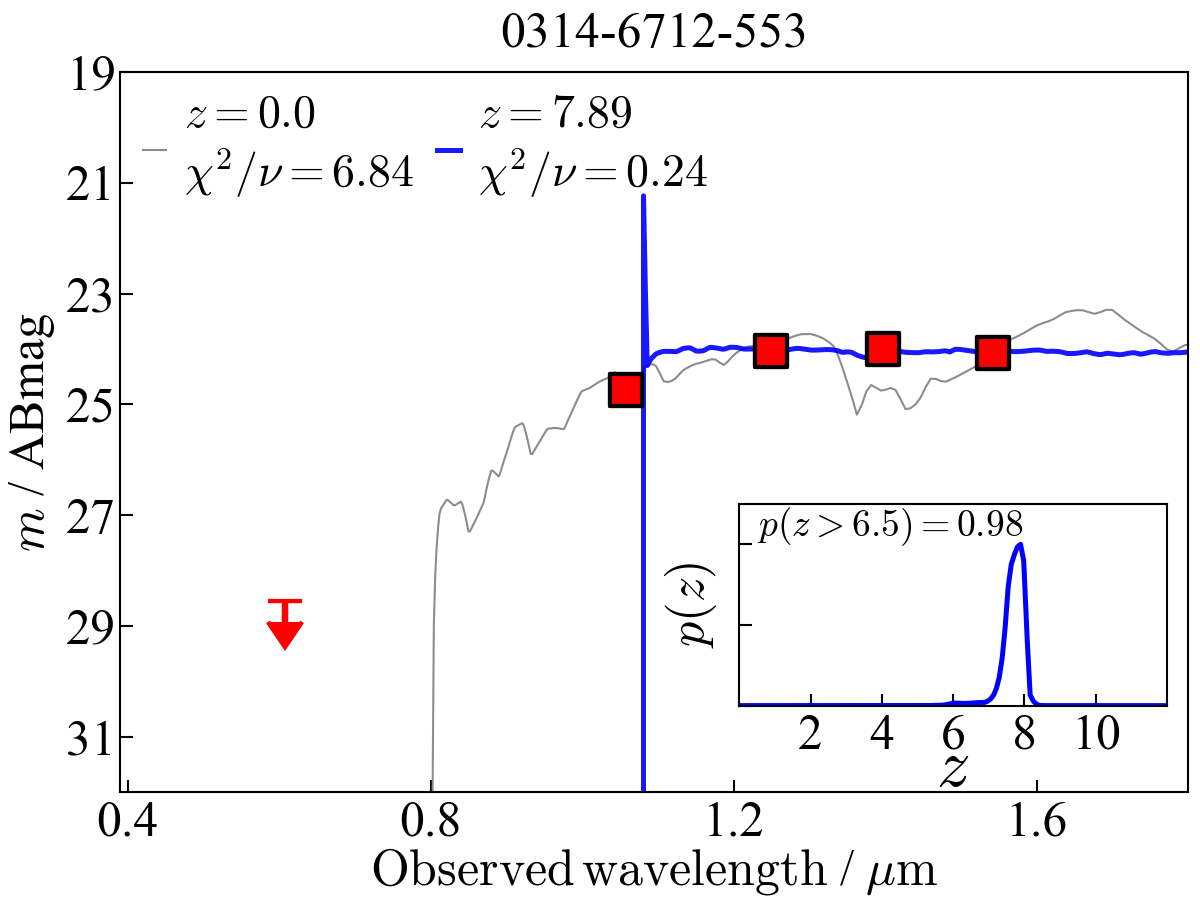}
	\includegraphics[width=0.33\textwidth]{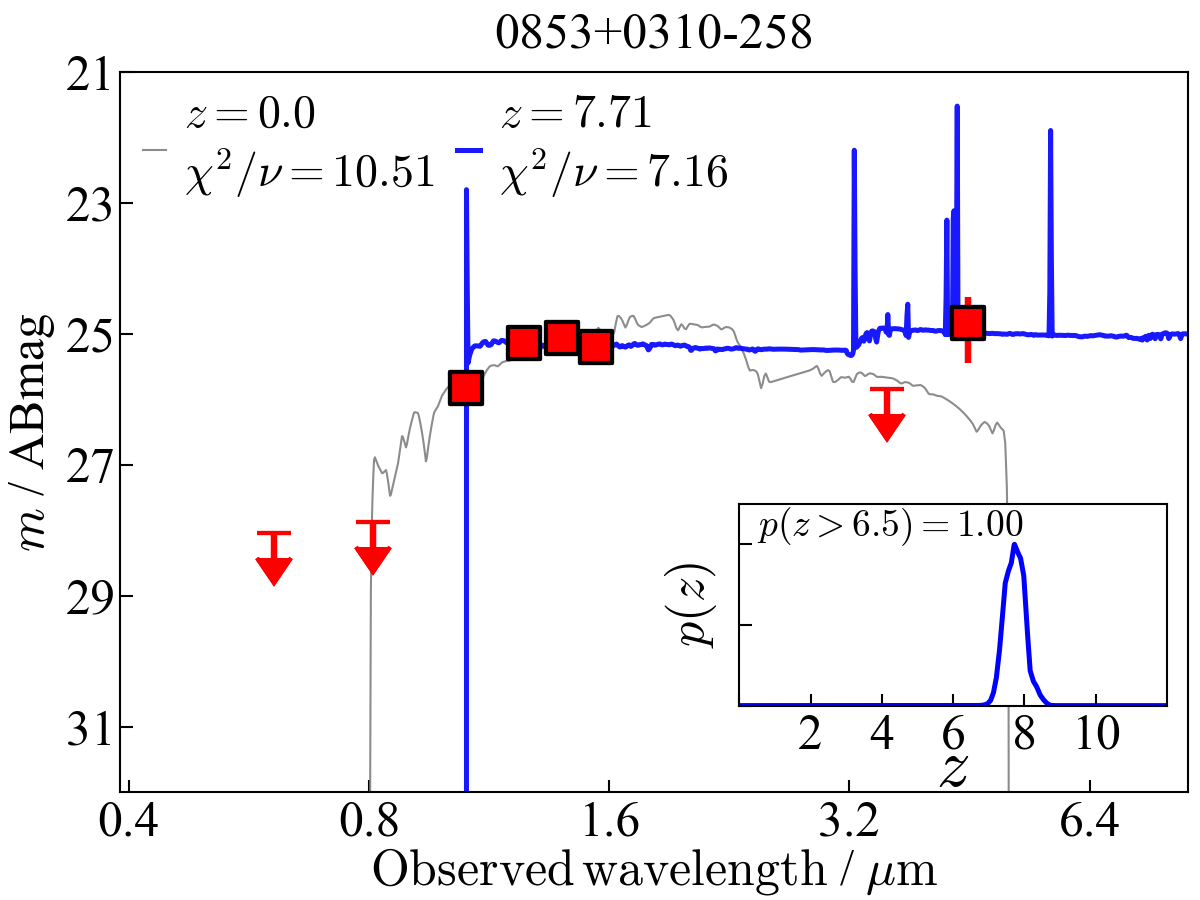}
	\includegraphics[width=0.33\textwidth]{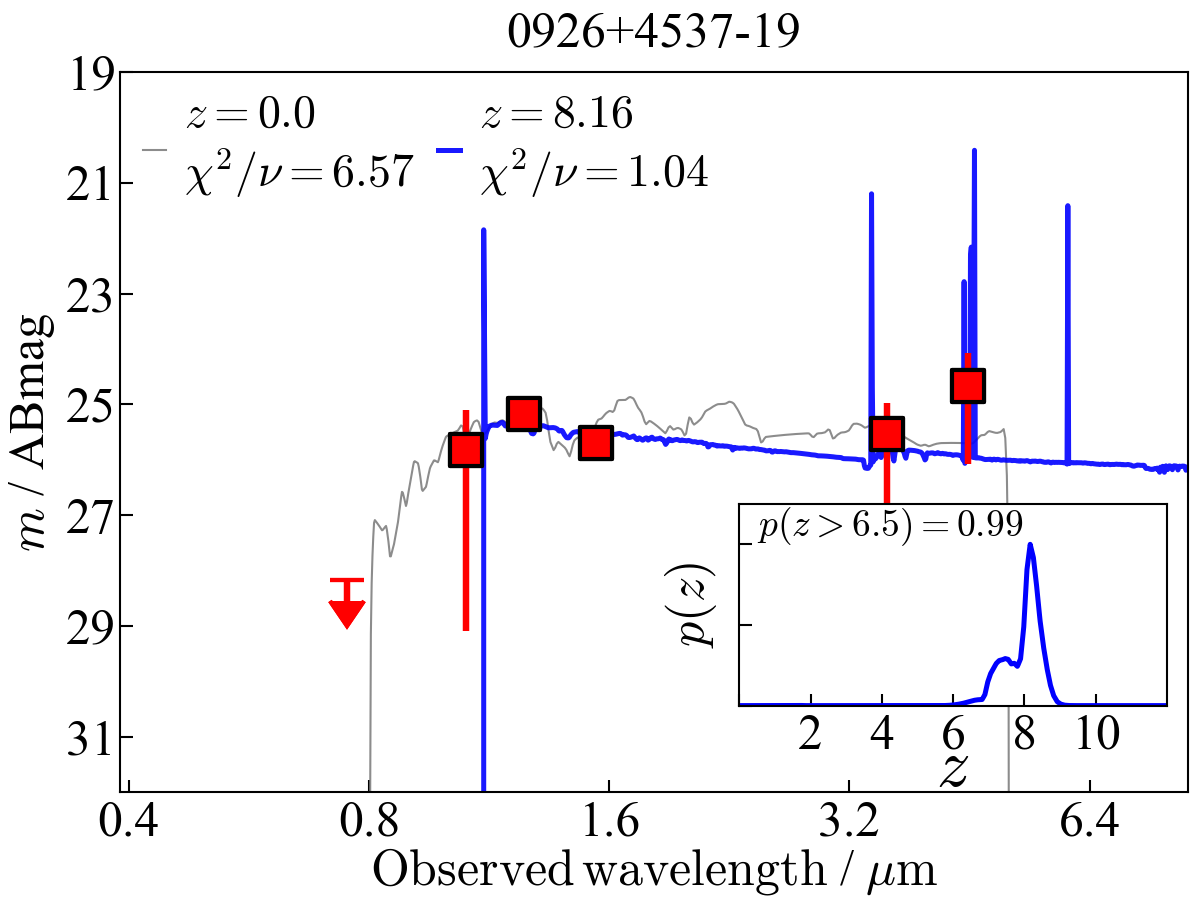}
	\caption{
	Spectral energy distributions (SEDs) of point sources from the z8\_Y105 selection, with photometric redshift probability distribution in inset. Observed fluxes (red squares for detection, and arrows for $1\,\sigma$ upper limits) and the best-fit templates at the peak redshift (blue solid lines) are shown. \yamato\ (middle) shows flux excess in ch2, likely from strong \hb+\oiii\ emission, resulting in a large reduced chi-square value (see also Sec~\ref{ssec:burst} and Fig.~\ref{fig:sed}). The best-fit dwarf templates are also shown (thin gray lines). It is noted that the dwarf templates used here are only available at $0.8<\lambda/\,\mu{\rm m}<5.5$, and flux at a shorter wavelength range is set to zero (Sec.~\ref{ssec:spec}). 
	}
\label{fig:sed_z8Y105}
\end{figure*}

\subsection{Selection of high-redshift candidates with point-source morphology}\label{ssec:sele}
We first select $z\sim8$ source candidates by the Lyman break dropout technique \citep{steidel96}, using {an updated version of} color-cut criteria presented in \citet[][hereafter ``z8\_Y105" selection]{calvi16};
$$S/N_{\rm non-detection\ filters}<1.0$$
$$S/N_{\rm 125}>6.0$$
$$S/N_{\rm 160}>4.0$$
$$Y_{105}-J_{125}>{0.6}$$
$$J_{125}-H_{160}<0.5$$
$$Y_{105}-J_{125}>1.5\cdot(J_{125}-H_{160})+{0.6}$$
where {we adopt a stricter color limit for Lyman break (cf. $Y_{105}-J_{125}>0.45$ in \citealt{calvi16})}. Non-detection filters are those bluer than F105W.

To further refine the selection, we exclude dropout candidates with $p(z)<0.7$, where $p(z)$ is photometric redshift probability at $z>6.5$, estimated by a photometric redshift code, \eazy\ \citep[][see Sec~\ref{ssec:phot} for details]{brammer08}. \numall\  source candidates are selected from 206\,fields where the z8\_Y105 selection is available.

We then refine this high-$z$ candidate catalog to extract objects with point-source morphologies. To configure optimal parameters for this, we conducted a source parameter recovery test, by adding point-source objects and extended objects to a set of real images, and then measured their photometric properties with \sext\ in the same way as for real sources. We follow \citet{morishita18b} to model point and extended sources at $z\sim8$, by assigning source radius, S\'ersic index, and UV slope. 

We include parameters such as size, elongation (ratio of major to minor axis radius, $e=a/b$), flux concentration, and CLASS\_STAR in the test. Among these parameters, the flux concentration parameter performs significantly better than other parameters in separating the two populations, up to $\sim26$\,mag (approximately the typical limiting magnitude of \spbg). Furthermore, we also set an upper limit in elongation to further improve the selection, although this is only a necessary but not a sufficient condition, as extended source with round shape would still pass the selection. We note that while CLASS\_STAR is often used as a star/galaxy separation indicator, our test indicates that its performance becomes less effective from a relatively bright magnitude, $\sim23$\,mag, as was previously reported \citep[e.g.,][]{finkelstein15}.

In conclusion, we set the following criteria to separate point sources from extended objects;
$$f_4/f_8>0.5$$
$$e < 1.2$$
where $f_{x}$ represents flux measured within a $r=x$\,pixel radius aperture ($0.\!\arcsec08$/pixel for \spbg). These parameters are measured in the detection band (F140W+F160W stacked images if both filters are available, and F160W for fields where F140W is not available). With these criteria, we find \num\ point sources from the z8\_Y105 sample. The postage stamps of these point sources are shown in Fig.~\ref{fig:stamp_z8Y105}.

\section{Nature of the point sources}\label{sec:nature}

\subsection{Photometric properties}\label{ssec:phot}
The selected point sources are shown in a size-magnitude diagram in Fig.\ref{fig:Mr}, and compared to other extended sources selected as z8\_Y105. We use the half-light radius measured by \sext\ for size here, and the absolute UV magnitude from the photometric redshift analysis below. By measuring the apparent size of stars at $\simlt24$\,mag among all \spbg\ fields, we found that $0.\!\arcsec2$ is a robust lower limit independent of source magnitude. Measurements smaller than this size are replaced by the lower limit in the figure and considered unresolved.

Of note is that \sext\ returns sizes smaller than this limit for all the point sources we selected with our criteria, and for three additional sources that do not meet them. Two of them are at the faint-flux end, where morphology measurement are less reliable due to low signal-to-noise ratio. The other one shows a faint elongated structure, and \sext\ measures $e=1.3$.

Fig.~\ref{fig:Mr} shows the star formation rate surface density, inferred from size and luminosity using the relation between UV magnitude and star formation rate \citep{kennicutt98,ono13,holwerda15}:
\begin{equation}
    M_{UV} = - 2.5  \log [ {\pi r^2 \cdot \Sigma_{\rm SFR} \over{2.8\times10^{-28} {\rm (M_\odot yr^{-1}) } } } ] +51.59.
\end{equation}
Given the resolution limit, the star formation rate surface density we infer for our point sources is a {\it lower} limit. Nevertheless, this lower limit is already in the range of starburst galaxies found in the local universe, $\sim1$-$100\,M_\odot\,{\rm yr^{-1} kpc^{-2}}$ \citep{kennicutt98}, implying high star formation rates. {It is noted that when these point sources are considered as quasars, the relation between star formation and absolute magnitude is not necessarily accurate, due to possible contribution from active galactic nuclei.}

\begin{figure}
\centering
	\includegraphics[width=0.45\textwidth]{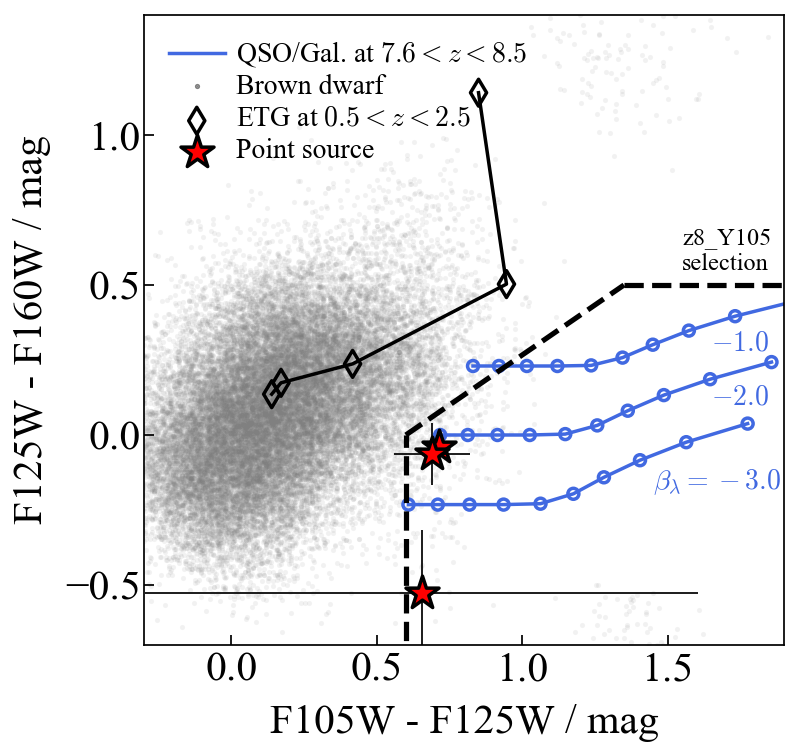}
	\caption{Color-color diagram for the z8\_Y105 selection (region bounded by dashed lines at right bottom). The three point sources of this study are shown (red stars). Colors of various populations are shown: high-$z$ objects with different UV slopes ($\beta_\lambda$), brown dwarfs taken from the IRTF spectral Library (gray dots), and early-type galaxies (diamonds). It is noted that scattered data points of brown dwarfs are most likely rejected by a photometric redshift analysis (Sec.~\ref{sssec:photz}).
	}
\label{fig:cc}
\end{figure}

\subsubsection{Photometric Redshift Analysis}\label{sssec:photz}
We use \eazy\ \citep{brammer08} with the default setup (v.1.3 templates), to derive redshift posterior probability and the best fit spectral energy distributions (SEDs). {We do not include templates for AGNs, as our primary goal here is to exclude low-$z$ contaminants with significant redshift probability at low redshift. Adding such templates may increase high-$z$ probability but not the opposite as the primary low-$z$ contaminants are those with red spectral features (see also below).}
We turned off the magnitude prior functionality of \eazy, as we focus on rare objects and they may not follow an empirical relation designed for galaxies. The posterior is therefore simply $\propto \exp \left[ -0.5 \chi^2/{\nu} \right]$, where $\chi^2/{\nu}$ is reduced chi-square. The results are shown in Fig.~\ref{fig:sed_z8Y105}. 

Two of the point sources (0853+0310 and 0926+4537) have IRAC coverage. Adding IRAC photometry significantly improves the photometric redshift estimate, by differentiating real high-$z$ candidates from low-$z$ interlopers. Fluxes in IRAC images are extracted by applying high-resolution (HR) morphological information as prior knowledge. We use \galfit\ \citep{peng02,peng10} on the detection band to get structural parameters (effective radius, S\'ersic index, and axis ratio), and fix these parameters for the IRAC image fitting, while leaving magnitude and position as free parameters. To account for flux contribution from nearby sources, we simultaneously fit sources brighter than 80\,\% of the target source flux within a radius of $8\arcsec$, while other sources detected in high resolution image are masked during the fit. Measured fluxes are summarized in Table~\ref{tab:mag}.

\begin{deluxetable*}{lccccccccc}
\tabletypesize{\footnotesize}
\tabcolsep=8pt
\tablecolumns{10}
\tablewidth{0pt} 
\tablecaption{Photometric fluxes of point sources in the z8\_Y105 selection.}
\tablehead{
\colhead{ID} & \colhead{F350LP} & \colhead{F606LP} & \colhead{F814W} & \colhead{F105W} & \colhead{F125W}  & \colhead{F140W}  & \colhead{F160W}  & \colhead{ch1} & \colhead{ch2}
\vspace{-0.3cm}\\
\colhead{} & \colhead{$\mu$Jy} & \colhead{$\mu$Jy} & \colhead{$\mu$Jy} & \colhead{$\mu$Jy} & \colhead{$\mu$Jy} & \colhead{$\mu$Jy} & \colhead{$\mu$Jy} & \colhead{$\mu$Jy} & \colhead{$\mu$Jy}
}
\startdata
0314-6712-553 & $<0.01$ & --- & --- & $0.46\pm0.01$ & $0.88\pm0.01$ & $0.90\pm0.01$ & $0.85\pm0.01$ & --- & --- \\
0853+0310-258 & $<0.02$ & --- & $<0.03$ & $0.17\pm0.02$ & $0.32\pm0.02$ & $0.34\pm0.02$ & $0.30\pm0.02$ & $<0.17$ & $0.42\pm0.18$ \\
0926+4537-19 & --- & $<0.02$ & --- & $0.17\pm0.16$ & $0.31\pm0.02$ & --- & $0.19\pm0.04$ & $0.22\pm0.15$ & $0.49\pm0.36$
\enddata
\tablecomments{
$1\sigma$ error are quoted for those with $S/N>1$, and $1\sigma$ upper limits for the rest of data points.
}
\label{tab:mag}
\end{deluxetable*}

The best-fit SED templates are shown in Fig.~\ref{fig:sed_z8Y105}. A strong color break and blue rest-UV slope are observed for all point sources, which make their redshift probability strongly peaked at $z\simgt8$. In fact, a blue UV slope is a key discriminant between high-$z$ objects from low-$z$ interlopers, as a Lyman break can be mimicked by a strong Balmer break of evolved populations (i.e. early-type galaxies) at $z\sim2$, dust attenuation of $z\simgt5$ galaxies, and brown dwarfs. Recent observations of $z\simgt7$ galaxies have revealed that only a fraction of those galaxies have a significant amount of dust \citep[][but see also \citealt{watson15,laporte17,tamura18} for detection of dust continuum]{akinoue16,hashimoto18}. By requiring a blue UV slope our selection is designed to be conservative and robust towards interlopers.

One of the point sources, \yamato, shows a significant flux excess in IRAC ch2, $[3.6]-[4.5]\simgt1.0$\,mag. The photometric redshift result therefore returns a large chi-square value (but see Sec.~\ref{ssec:burst} for more sophisticated fitting results). In combination with the non-detection in the deep ch1 image (7200\,s, with $1\,\sigma$ limiting magnitude of $25.7$\,mag), the excess implies a strong \hb+\oiii\ emission at the peak photometric redshift. This object was first identified by \citet{calvi16}, and also presented in \citet{bridge19} with a \hst\ F814W follow-up image, as a promising $z\sim8$ candidate. However, the full-depth ch2 image was not available at the time of these studies ($\sim900$\,s; whereas 7200\,s here), and neither of them revealed the flux excess. \citet{bridge19} reported a weak flux in ch1 ($\sim25.4$\,mag), but not a significant detection (S/N $\sim1.1$). The ch1 image analyzed here shows a small portion of positive pixels in ch1 (Fig.~\ref{fig:stamp_z8Y105}), but the extracted total flux is not significant compared with the noise estimated from the image. The difference may be attributed to differences in the flux extraction procedure. The results reported here do not vary if the upper limit or the low S/N detection are used. Our spectroscopic follow-up of this object is presented in Sec.~\ref{ssec:spec}, and photometric properties are discussed in more detail in Sec.~\ref{ssec:burst}.

{In Fig.~\ref{fig:cc}, we show a color-color diagram for the z8\_Y105 selection, along with colors of high-$z$ sources and other possible low-$z$ interlopers. For high-$z$ galaxies and quasars, we generate spectral templates by assuming a single slope for different UV $\beta$ slopes ($\beta_{\lambda}=-1,-2$, and $-3$). We do not include emission lines, such as \ly, as is often expected for starburst galaxies and quasars. Adding such a component would change the colors but toward right bottom in the plot, making the colors even more separable from the other contaminating populations. Colors of brown dwarfs are calculated by using spectra taken from the IRTF spectral Library \citep{rayner03}. Their colors are scattered for 100 times by applying a random photometric error for \spbg\ ($\sim0.1$\,mag). Lastly, colors of early-type galaxies are calculated based on synthetic templates generated by \fsps\ \citep{conroy09fsps}, with a simple population of $1$\,Gyr old without dust attenuation.}

While low-$z$ galaxies can be excluded from the point-source selection based on morphology, it is still possible that cool stars, primarily T/L/M types (i.e. brown dwarfs), are selected in the z8\_Y105 selection due to photometric scatter. {As is seen in the color-color diagram, while brown dwarfs dominate much bluer region than our z8\_Y105 population,} there is a non-trivial fraction of cool stars that migrate into the selection due to photometric scatter ($\sim1.2\%$, assuming the same photometric error as above). To further assess the possibility of such contamination by foreground stars, we repeat the phot-$z$ fitting process with dwarf templates. A set of dwarf templates is taken from the IRTF spectral Library, provided to \eazy, and fit to the data with redshift fixed to 0. The fits (shown in Fig.~\ref{fig:sed_z8Y105}) result in large $\chi^2/\nu$ values for all three point sources (Table~\ref{tab:tab2}). Template mismatch primarily occurs at $1.5\,\mu$m, as well at the wavelength of IRAC bands when available. It is noted that the dwarf templates used here only cover $0.8<\lambda/\,\mu{\rm m}<5.5$. Due to this artificial truncation at $<0.8\,\mu$m \citep[see, e.g.,][]{pirzkal05}, the $\chi^2/\nu$ values could become even larger with a dwarf spectrum of complete wavelength coverage.

\begin{figure*}
\centering
	\includegraphics[width=0.99\textwidth]{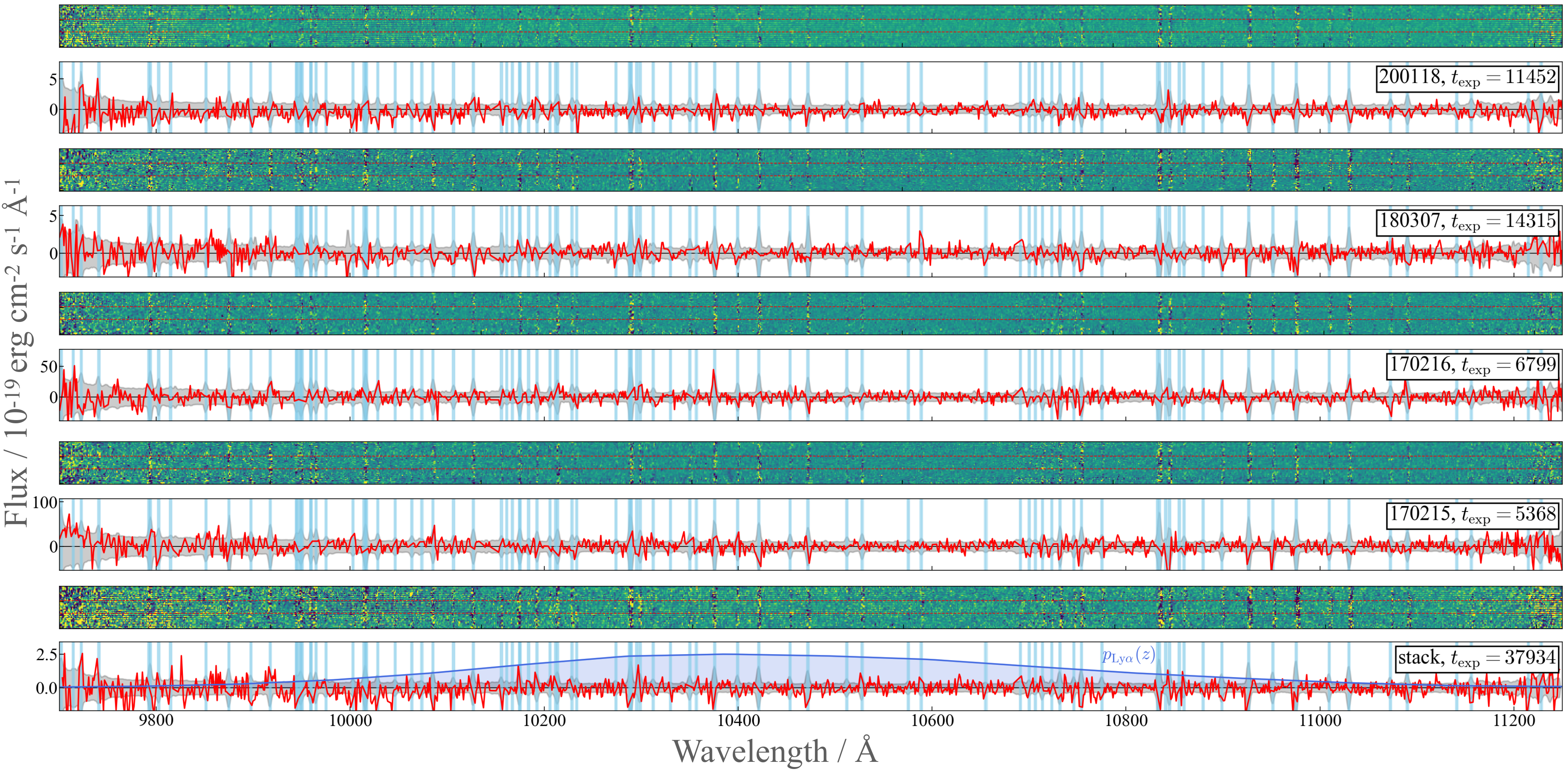}
	\caption{
	Two-dimensional spectra of \yamato\ taken with Keck/MOSFIRE ($Y$-grating) on 4 nights (January 2020, March 2018, and two nights from January 2016, from top to bottom), as well as all-stacked spectrum. Optimally extracted one-dimensional spectra are also shown (red solid lines, and gray shaded region for 1\,$\sigma$ uncertainty), with sky lines being masked (light blue stripes). No \ly\ emission is revealed in the spectra. {In the bottom panel, we show the probability distribution of \ly\ (blue shaded region) from the photometric analysis (Fig.~\ref{fig:sed_z8Y105}). The distribution is normalized arbitrarily.}
	}
\label{fig:2d}
\end{figure*}


\begin{deluxetable*}{lcccccccccc}
\tabletypesize{\footnotesize}
\tabcolsep=10pt
\tablecolumns{22}
\tablewidth{0pt}
\tablecaption{Photometric properties of point sources in the z8\_Y105 selection.}
\tablehead{
\colhead{Field ID} & \colhead{Survey} & \colhead{ObjID} & \colhead{R.A.} & \colhead{Decl.} & \colhead{$z_{\rm peak}$}  & \colhead{$M_{\rm UV}$}  & \colhead{$r$}  & \colhead{$m_{125}$} & \colhead{$\chi^2/\nu$} &  \colhead{${\chi^2/\nu}_{\rm dw}$}
\vspace{-0.3cm}\\
\colhead{} & \colhead{} & \colhead{} & \colhead{deg} & \colhead{deg} & \colhead{}  & \colhead{mag}  & \colhead{kpc}  & \colhead{mag} & \colhead{} & \colhead{}
}
\startdata
0314-6712 & BoRG cycle22 & 553 & 48.449680 & -67.209724 & 7.89 & -23.2 & 0.96 & 24.1 & 0.24 & 6.84 \\
0853+0310 & BoRG cycle22 & 258 & 133.185590 & 3.146692 & 7.71 & -22.0 & 0.97 & 25.3 & 7.16$^\dagger$ & 10.51 \\
0926+4537 & HIPPIES cycle18 & 19 & 141.586580 & 45.593613 & 8.16 & -21.8 & 0.94 & 25.2 & 1.04 & 6.57
\enddata
\tablecomments{
$z_{\rm peak}$ : Photometric redshift estimated with \eazy.
$\chi^2/\nu$ : Reduced chi-square from \eazy\ photometric redshift fitting analysis with extra-galactic templates.
$\chi_{\rm dw}^2/\nu$ : Reduced chi-square with brown dwarf templates.
$\dagger$ : Fitting result including IRAC ch2 data point. Table~\ref{tab:sed} shows the best-fit result after excluding the data point.
}
\label{tab:tab2}
\end{deluxetable*}

\subsubsection{Astrometric Analysis}\label{ssec:astro}
To further investigate possible contamination by brown dwarfs, we measure astrometry of one of the point sources, \yamato, as the field has two separate images of F140W taken in 2010 and 2015. The former was taken as a part of direct images for a pure-parallel grism survey \citep[WISP;][]{atek10}, and thus is relatively shallow ($\sim250$\,sec), while the latter was taken in BoRG cycle22 ($\sim$1300\,sec). 

We reduce these data separately in the same way as described above. Absolute astrometry was not calibrated for the 2010 image. We, instead, measure the relative distance of the target from the light-weighted center of each image. The coordinate of light-weighted center is calculated using the same set of 35 bright extended sources within a $<1\arcmin$ radius around the target in each image. Displacement between the light-weighted center and \yamato\ is calculated in each image, and used to infer any relative motion of the target over the two epochs.

The inferred motion is $\Delta({\rm RA}, {\rm Dec}) = (0.009 \pm 0.040, 0.142 \pm 0.140)$ in arcsec, {where the associated uncertainty is the median of the displacements calculated for bright, extended sources (i.e. extragalactic sources) in the same field. For this error calculation, we identify 50 common sources in both images, and calculate displacement of each source in the same way as for \yamato.} The relatively large uncertainty in declination may be attributed to image alignment with a small number of sources available in the image taken in 2010. Our conclusion is therefore that the observed shift is not significant, and \yamato\ is, in the first place, not a foreground star with significant motion, further securing the conclusion in Sec.~\ref{ssec:phot}. A $\sim1.5$\,mag deeper image will be required to detect any motion of \yamato\ at sub-arcsec precision \citep{su11}.

\subsection{Spectroscopic follow-up}\label{ssec:spec}
Two Keck nights were allocated to observe the field including \yamato\ in January 2020 with MOSFIRE (S19B, PI T. Morishita/ UC2019B, PI T. Treu). The first night was cancelled due to bad weather, and the second night was executed under partially cloudy condition with some fog, with average seeing $\sim0.\!\arcsec9$. The data were taken with the $Y$-grating and ABA'B' dithering. We reduced the data by using a pipeline developed by the MOSDEF team \citep{kriek15}, which allows a differential-weighting stacking depending on the atmospheric seeing size, measured by a star assigned in one of the slit masks. After removing data with seeing $>1$\,arcsec, the total exposure time is $\sim3.2$\,hr. 

In addition to our observing runs, we retrieved archival data of the same object from the Keck Observatory Archive --- two nights from January 2016 (U092; PI G. Illingworth) and one night from March 2018 (N101; PI J. Bridge). The data sets from 2016 have 1.5\,hr and 1.9\,hr on-source exposures, with average seeing condition of $\sim0.\!\arcsec8$ and $0.\!\arcsec75$, respectively, measured from one of the monitoring stars. The data from 2018 were taken under good weather condition (J. Bridge, private communication), though the seeing during the night is unknown as no monitoring star was included in the configuration mask. We therefore give all frames equal weight, resulting in $4.0$\,hr exposure for this night. We reduced these data in the same way as for our own data. 

Rectified two-dimensional spectra from each night are shown in Figure~\ref{fig:2d}. \ly\ emission is not detected over the entire wavelength range. {Furthermore, we stack these reduced two-dimensional spectra by taking weighted average (bottom panel). The stacked spectrum still does not reveal \ly\ emission.}

{By using the stacked spectrum, we then aim to obtain an upper limit for \ly\ emission.} Following \citet{hoag19}, we estimate the limiting flux by
$$f_{\rm lim}=\Delta \lambda \times \sqrt{ {2 {\rm FWHM_{\rm inst}}\over{\Delta \lambda} } } \sigma(\lambda)$$
where $\Delta \lambda$ is the pixel scale of MOSFIRE ($=1.086\,\AA$), FWHM$_{\rm inst}$ is the instrumental spectral resolution ($\sim3$\,\AA), and $\sigma(\lambda)$ is noise spectrum extracted from the stacked spectrum. 

This procedure gives us a $5\,\sigma$ limiting flux of {$\sim{\rm 7.8\times10^{-19}\,erg/s/cm^2}$}, from a $1\sigma$ probability range of \ly\ line calculated in the photometric redshift analysis (Fig.~\ref{fig:2d}), for an unresolved line. The flux limit corresponds to rest-frame \ly\ equivalent width of {13\,\AA$/(1+z)$}, assuming the continuum flux from its F125W magnitude ($\sim25.1$\,mag). The limit is sufficiently low to detect the extreme emission expected for low-luminosity quasars \citep{matsuoka19}, or luminous galaxies \citep[e.g.,][]{oesch15,zitrin15,song16}. Our non-detection of \ly\ therefore implies that this object is possibly located in intergalactic medium (IGM) of significantly high neutral fraction, which most of \ly\ photons cannot escape from, or that \ly\ line is significantly broader than those of the previously reported luminous objects. It is noted that the limiting flux scales with $\sqrt{\rm FWHM/FWHM_{inst}}$ for a resolved line. However, at this luminosity ($M_{\rm UV}\sim-22$\,mag) line width of \ly\ is not necessarily broad even for low-luminosity quasars at $\sim6$ \citep[e.g., HSC J2228+0128 in][]{matsuoka16}. With this caveat noted, we discuss the spectroscopic non-detection in the following section.

\begin{figure*}
\centering
	\includegraphics[width=0.9\textwidth]{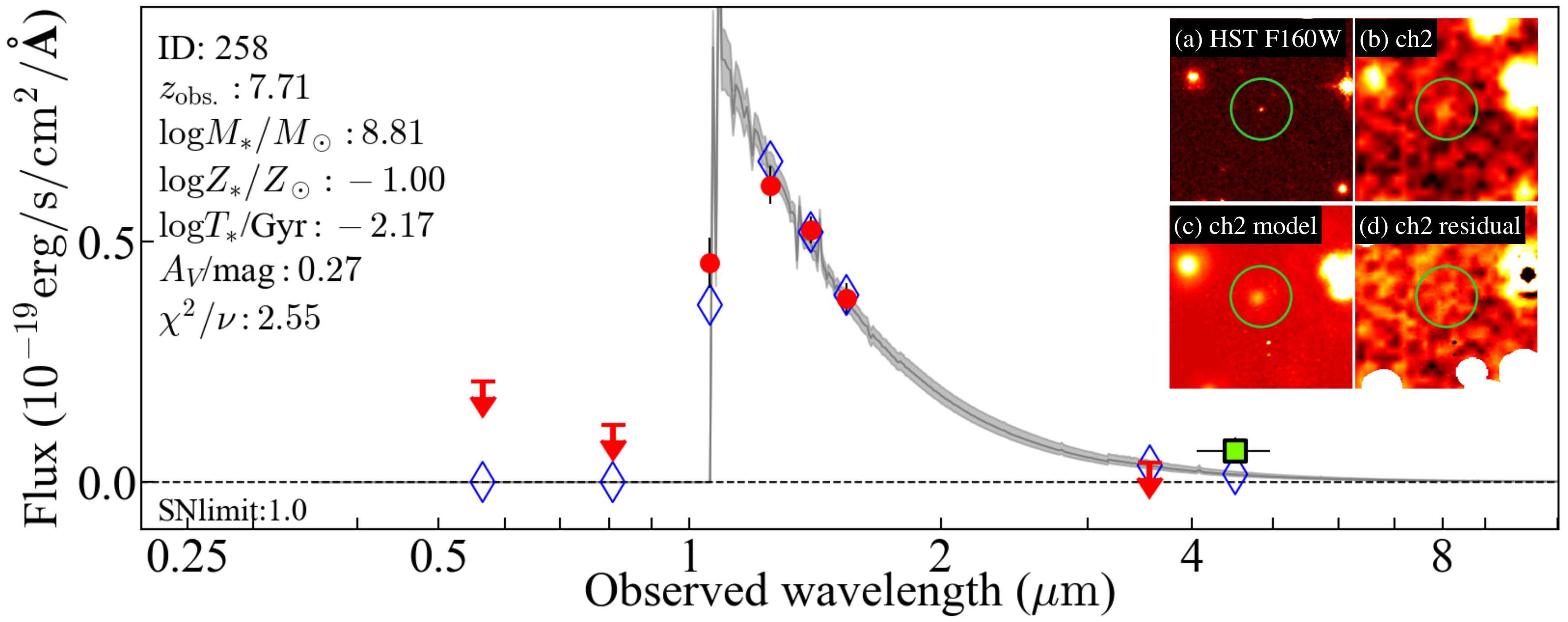}
    \caption{
	Best-fit SED fitting results of \yamato, with \fsps, $\logZ=-1$. Broadband photometric data points (red circles for detection, and triangles for non-detection) are fitted, while the flux excess in IRAC ch2 (green square symbol) is excluded. The best fitted template, with 16/84th percentile range, is shown (gray line and shaded region), with its expected broadband fluxes (blue diamonds).
	In inset, observed \hst's detection image (a), observed IRAC ch2 image (b), model (c), and residual image (d) are shown. Regions masked during fitting are left blank in the residual image. 
	}
\label{fig:sed}
\end{figure*}

\section{Discussion}\label{sec:lum}

\subsection{Physical properties of \yamato}\label{ssec:burst}

The ch2 excess points towards an unusual nature of \yamato\. To further investigate its nature, we analyze its photometry with \gsf\ \citep[v1.3;][]{morishita18,morishita19}, a Bayesian SED fitting code. We provide a set of composite stellar population templates generated by \fsps\ \citep{conroy10}, setting the initial mass function to \citet{chabrier03}. Repeating the analysis with a set of templates from B-PASS v.2.2 \citep{eldridge17,stanway18} does not change our conclusions. {Our analysis here is based on an assumption that the object can be well represented by stellar spectrum; distinction between luminous galaxies and low-luminosity quasars at this redshift may rather be ambiguous, and this is beyond the scope here as such distinction cannot be meaningfully constrained by our data presented here.}

To robustly measure the underlying continuum, we exclude the IRAC ch2 data point from the fitting. The IRAC ch1 band covers rest-frame $\sim4000$\,\AA, and the non-detection implies that there is no significant emission from the \oii3726+3728\,\AA\ doublet. Due to the limited number of photometric data points (4 detections excluding IRAC ch2, and 3 non-detections), we separately run the fit fixing metallicity to either solar or 10\%\ solar values. Results from each fit are summarized in Table~\ref{tab:sed}. Redshift is fixed to the peak of the posterior derived by \eazy.

\begin{deluxetable*}{cccccccccc}
\tabletypesize{\footnotesize}
\tabcolsep=11pt
\tablecolumns{10}
\tablewidth{0pt} 
\tablecaption{Spectral energy distribution fitting results of \yamato.}
\tablehead{
\colhead{$z$} & \colhead{$M_*$} & \colhead{SFR} & \colhead{$Z_*$} & \colhead{$T_*$} & \colhead{$A_V$}  & \colhead{$M_{\rm UV}$} & \colhead{$\chi^2/\nu$} & \colhead{$EW_{\rm 0,ch2}$} & \colhead{$f_{\rm esc,Ly\alpha}$}
\vspace{-0.3cm}\\
\colhead{} & \colhead{$\log M_\odot$} & \colhead{$\log M_\odot {\rm yr^{-1}}$} & \colhead{$\log Z_\odot$} & \colhead{$\log$\,Myr} & \colhead{mag} & \colhead{mag} & \colhead{} & \colhead{\AA} & \colhead{\%}
}
\startdata
$[7.71]$ & $8.8_{-0.2}^{+0.3}$ & $1.0_{-0.2}^{+0.2}$ & [$-1.0$] & $0.8_{-0.5}^{+0.5}$ & $0.3_{-0.2}^{+0.2}$ & $-22.1_{-0.0}^{+0.0}$ & $2.55$ & $2961\pm{1666}_{-243}^{+396}$ & $<1.7$\\
$[7.71]$ & $8.7_{-0.2}^{+0.3}$ & $0.8_{-0.2}^{+0.3}$ & [$0.0$] & $0.8_{-0.5}^{+0.5}$ & $0.2_{-0.1}^{+0.2}$ & $-22.0_{-0.0}^{+0.0}$ & $2.68$ & $2460\pm{1457}_{-183}^{+381}$ & $<2.0$
\enddata
\tablecomments{
Parameters with values bracketed are fixed during fit. Two values for metallicity ($Z_*$) are examined. SFR : Averaged star formation rate measured within the last 30Myr of the marginalized star formation history. $EW_{\rm 0,ch2}$ : Rest-frame equivalent width measured with IRAC ch2 excess from the marginalized continuum. Associated uncertainties represent 1-$\sigma$ random error from the ch2 flux estimate and 16/84th percentiles range from the posterior SED.
}
\label{tab:sed}
\end{deluxetable*}

The best-fit result with $\logZ=-1$ ($\chi^2/\nu=2.55$) is shown in Fig.~\ref{fig:sed}. While the slight preference toward low metallicity is reasonable given the young age of the universe at this redshift, the fit is not statistically significant better than that with solar metallicity ($\chi^2/\nu=2.68$). It is challenging to determine metallicity only by optical broadband photometry \citep[e.g.,][]{morishita19}; the inferred metallicity may be attributed to the age-metallicity-dust degeneracy as well as assumptions in star-formation history, and therefore we do not discuss physical interpretation of the inferred metallicity here.

From the best-fit spectrum, the rest-frame equivalent width of \hb+\oiii\ lines can be calculated as;
\begin{equation}
    EW_{\rm 0,H\beta+[OIII]} = {(f_{\rm ch2} - f_{\rm cont}) \over{f_{\rm cont}}} {\Delta\lambda_{\rm ch2} \over{ (1+z)}}
\end{equation}
where $f_{\rm cont}$ is the underlying continuum flux from the posterior SED, $f_{\rm ch2}$ is the observed ch2 flux, and $\Delta\lambda_{\rm ch2}$ is the full-width half maximum of the IRAC ch2 filter. 

Our estimate from the best fit is $EW_{\rm 0,H\beta+[OIII]} \sim 3000\pm{1700}_{-240}^{+400}$\,\AA, where the first uncertainty refers to the flux error in ch2, and the second to the uncertainty in the continuum estimate (16/84th percentiles). The inferred equivalent width is at the high end of those of $z\simgt7$ galaxies estimated from IRAC excess \citep{labbe13,smit14,smit15,roberts-borsani16,debarros19}. Such a high equivalent width value is not physically impossible. Based on a stellar population+nebular emission model \citep[see Section 3.2 of][]{oesch07}, equivalent width can reach up to $\sim15000$\,\AA\ at moderate oxygen abundance, high gas temperature, and young stellar population \citep[see also][who spectroscopically measure $EW_{\oiii}\sim4400$\,\AA]{schenker13b}. In contrast, at low metallicity, contribution from \oiii\ becomes significantly less, and the maximum equivalent width reached is $\sim1500$\,\AA.

\subsubsection{Why no \ly?}\label{ssec:esc}
A critical question is, then, why we do not detect any \ly\ from such an extreme object. The non-detection of \ly\ and extremely high \hb+\oiii\ equivalent width can be attributed to two factors --- small \ly\ escape fraction, and large \oiii-to-\hb\ ratio. For typical galaxies at high redshift, where the neutral hydrogen fraction is significantly high, it is not unusual to find galaxies with extremely low \ly\ escape fraction \citep{hoag18,mason18}. However, this is not necessarily the case for luminous sources \citep[][]{hu16,matthee18,mason18b}, as they may be able to create a large ionizing bubble and increase the escape fraction \citep{cen00,tilvi20}.

The  \oiii-to-\hb\ ratio is primarily determined by a combination of temperature and metallicity of the circumgalactic medium. For example, \citet{panagia03} calculated photoionization models and showed that the ratio spans $\sim0.7$ (at temperature of $3\times10^4$\,K) to 10 ($10^5$\,K) over a metallicity range of 0.1 to 1\,$Z_\odot$; this could be as low as $\sim10^{-2}$ at extremely low metallicity ($10^{-4}$\,$Z_\odot$).

While it is challenging to determine the contribution of each factor without direct spectroscopic observations of \hb\ and \oiii\ lines, we can estimate an upper limit of the \ly\ escape fraction, under the reasonable approximation of the relation between \ly\ and \hb\ luminosity for Case B recombination with a temperature of $10^4$\,K and an electron density of  $n_e=350$\,cm$^{-3}$ \citep{sobral19};
\begin{equation}
f_{\rm esc, Ly\alpha} = {L_{\rm Ly\alpha} \over{8.7L_{\rm H\alpha} \cdot 10^{0.4A_{\rm H\alpha} } } }.
\end{equation}

We use the measured limit on equivalent widths of  {\ly\ (13\,\AA\,$/(1+z)$)} and \hb+\oiii\ to calculate their luminosity, and \oiii-to-\hb\ ratio of $\sim10$, and other fiducial assumptions, such as \oiii\ 5007\,\AA-to-4959\,\AA\ ratio \citep[$\sim3$;][]{dimitrijevic07}, Balmer decrement ($\sim2.86$, under Case B), and dust attenuation ($A_V\sim0.3$) from the SED fitting analysis above. This procedure yields an upper limit to the \ly\ escape fraction of $f_{\rm esc, Ly\alpha}\sim \fesc \%$. It is noted that changing any of assumptions makes the estimated upper limit even smaller.

All things considered, we conclude that this luminous object is likely located in a significantly neutral region, but still with high temperature and moderate oxygen abundance that can boost \oiii\ (rather than \oii, which we do not see in ch1) emission. This is consistent with an increasing trend of \oiii/\oii\ with redshift as presented by \citet[][]{khostovan16}, which can be attributed to a trend in gas temperature and ionization parameter \citep[e.g.,][]{nakajima14}. It is unlikely that our target has very low oxygen abundance, since \hb\ by itself cannot account for the observed equivalent width. 

The physical interpretation of the low-\ly\ escape fraction is not as clear-cut. For example, \ly\ escape fraction can be affected by halo mass --- if a source resides in a massive halo, neutral hydrogen may be also ionized by surrounding sources \citep[][see also \citealt{dayal18}]{ren19,whitler20}. However, the non-negligible scatter in the halo mass-luminosity relation leaves a possibility that \yamato\ may be in a less-dense region, within a highly neutral intergalactic medium. 

As mentioned above, the limiting flux estimated here is for unresolved line, which scales with $\sqrt{\rm FWHM/FWHM_{\rm inst.}}$. It is thus still possible that \ly\ of this object has a much broader line profile, than previously reported values of luminous galaxies ($\simlt15$\,\AA), resulting in non-detection in our deep spectra. While a few spectroscopic results are available, our knowledge of {\it luminous} galaxies and {\it faint} quasars at this redshift is still limited, and it is challenging to expect the line profile solely from other photometric properties. A dedicated study on the distribution of \ly\ line width at this luminosity range would shed light on this uncertainty.

Spectroscopic confirmation of the source redshift will allow for decisive progress into understanding the nature of this source, by reducing many of the degeneracies in the interpretation and hopefully providing a line flux. UV lines such as \ciii\ \citep{stark17,Mainali17,schmidt17,laporte17b,hutchison19} may be sufficiently bright to be detected from ground-based facilities, while {JWST} will be able to observe rest-frame optical lines.

\begin{figure}
\centering
	\includegraphics[width=0.5\textwidth]{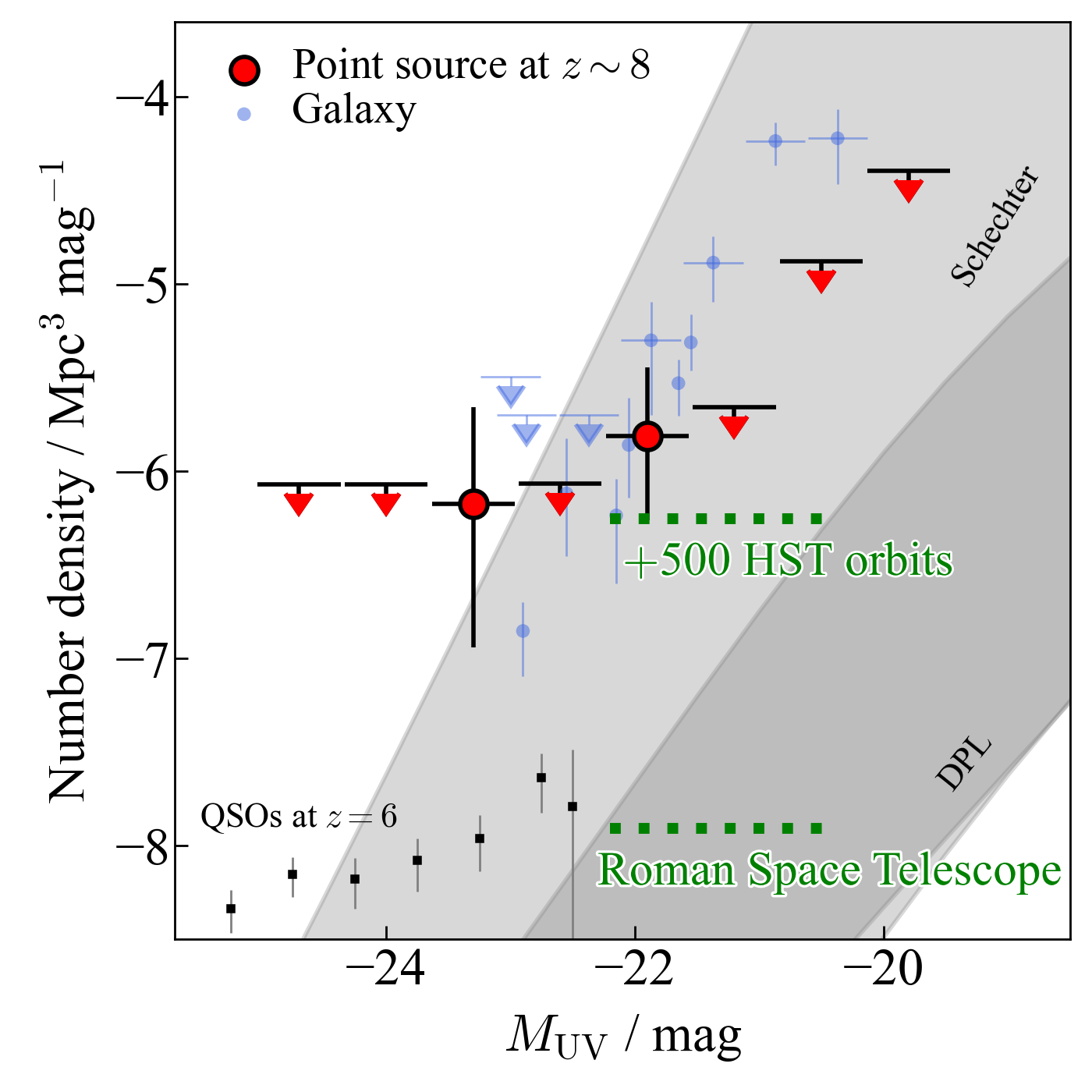}
	\caption{
	Number density of point sources at $z\sim8$ (red circles and arrows), compared with those of bright galaxy candidates at $z\sim8$ from other studies (blue circles and arrows). None of the point sources here are spectroscopically confirmed, and thus their number densities should be considered as upper limits for quasars'. Two shaded regions are empirical expectation for luminosity functions of quasars, assuming the Schechter (top) and double-power law (bottom) shapes \citep[][]{manti17}. Adding another cycle of BoRG ($\sim500$\,orbits) will improve the constraint at the low-luminosity range at $z\sim8$. To find a quasar of $M_{UV}\simlt-23$, {\it at least} $>100\times$ volume is required, as is suggested by $z\sim6$ results \citep[black squares;][]{matsuoka19}. {The Roman Space Telescope (RST)} can prove such a volume with a similar observing time.
	}
\label{fig:LF}
\end{figure}

\subsection{Number density of point sources at $z\sim8$}\label{ssec:nd}
In this last section, we calculate the number density of the point sources selected as $z\sim8$. We calculate the effective survey volume in a standard manner \citep{oesch12,carrasco18}, by adding artificial objects to images of each observed field and then investigated completeness of source identification through the same color selection method. Given our focus on point sources, the artificial objects are pure F160W point spread functions. The point-like nature of our sources results in a $\sim3\%$ ($30\%$) increase in effective volume compared to typical extended sources in the range $M_{\rm UV}=-24$ ($-22$). The estimated effective volume for the z8\_Y105 selection is $\sim2 \times10^{6}$\,Mpc$^3$ at $M_{\rm UV}\sim-24$\,mag and $\sim10^4$\,Mpc$^3$ at $M_{\rm UV}\sim-20$\,mag (Table~\ref{tab:tab1}).

\begin{deluxetable}{cccc}
\tabletypesize{\footnotesize}
\tabcolsep=12pt
\tablecolumns{4}
\tablewidth{0pt} 
\tablecaption{Number density of point sources at $z\sim8$}
\tablehead{\colhead{$M_{\rm UV}$} & \colhead{Volume$^{\rm a}$} & \colhead{$N_{\rm obj}$} & \colhead{Number density$^{\rm b}$}
\vspace{-0.3cm}\\
\colhead{(mag)} & \colhead{($10^3$\,Mpc$^{3}$)} & \colhead{} & \colhead{($\log$\,Mpc$^{-3}$ mag$^{-1}$)}
}
\startdata
$-24.7$ & 2146 & 0 & $<-6.07$\\
$-24.0$ & 2146 & 0 & $<-6.07$\\
$-23.3$ & 2146 & 1 & $-6.18_{-0.52}^{+0.77}$\\
$-22.6$ & 2127 & 0 & $<-6.07$\\
$-21.9$ & 1853 & 2 & $-5.81_{-0.36}^{+0.45}$\\
$-21.2$ & 832 & 0 & $<-5.66$\\
$-20.5$ & 138 & 0 & $<-4.88$\\
$-19.8$ & 45 & 0 & $<-4.39$
\enddata
\label{tab:tab1}
\tablenotetext{\rm {\bf Notes.}
}
{
\\
$\rm ^a$ Effective volume calculated by the completeness simulation.\\
$\rm ^b$ 1~$\sigma$ uncertainty calculated based on \citet{gehrels86} is quoted.
}
\end{deluxetable}
\
 
In Figure~\ref{fig:LF}, we show the calculated number density of point sources at $z\sim8$. Number densities of galaxy candidates, taken from previous studies \citep{bouwens15,livermore18,stefanon19,bowler20}, are also shown for comparison. The upper limits on the number density of point sources are already below the density of galaxies in the range $M_{\rm UV}=-22$ to $-21$\,mag.

{Due to the fact that none of the point sources presented in this study are confirmed as quasars, all data points in the plot should rather be considered as upper limits if taken as the number density for quasars. Despite the caveat, it is still worthwhile to mention that the volume probed by our survey is not sufficient to provide insight into possible evolution of the quasar luminosity function.} 

For example, two LFs extrapolated from low redshift are shown in the same figure---one with a Schechter form, and the other with a double-power law \citep{manti17}. Given the uncertainties represented by the gray bands, both extrapolations are consistent with the observations, although interestingly our upper limits are overlapping with the Schechter-based extrapolation, implying that our volume is approaching an interesting size at the faint end of our range. At the bright end, a comparison with the number densities of quasars at $z\sim6$ \citep{matsuoka19}, shows that we need approximately a factor of  $\sim100\times$ increase in volume to plausibly detect one or derive interesting limits.

Establishing the quasar luminosity function at these high redshift is extremely important to understand their seeding and growth mechanism given the short amount of time elapsed since the Big Bang. We show that sufficient increases in survey volume are well within reach of existing and planned space missions, which are essential to overcome the limitations of ground based surveys at these wavelengths.
In Fig.\ref{fig:LF} we show the volume density probed by adding further 500\,\hst\ orbits (similar to cycle22 BoRG, with $87$\,sightlines) and by the {Roman Space Telescope, RST}\footnote{https://roman.gsfc.nasa.gov} with a comparable observing time, assuming a similar sensitivity and filter combination \citep[Z087, Y106, J129, and H158;][]{rst20} as for our \hst survey. While continuing surveys like BoRG with \hst\ will still be beneficial for exploring low-luminosity quasars and luminous galaxies ($M_{\rm UV}\sim-22$), it is clear that {RST} will be a game-changer at the bright-end.

\section{Summary}\label{sec:sum}

We carried out a systematic search for quasars at $z\sim8$, using the \spbg\ data set, a compilation of \hst\ parallel observations from \numfld\ fields ($\sim$\,\Afld). This is to our knowledge the first dedicated search for point-like high redshift sources, since they were generally discarded as low redshift interlopers in previous studies. Our findings are summarized as follows;

\begin{enumerate}
 \item Based on the analysis of the spectral energy distribution, and astrometric analysis for one of the sources, we concluded that none of the three point sources selected are likely to be low-$z$ interlopers, including known types of brown dwarfs.

\item Our spectroscopic follow-up of \yamato\ did not reveal strong \ly\ emission down to a $5\sigma$ sensitivity of {$7.8\times10^{-19}$\,erg/s/cm$^2$} for an unresolved line.

\item The spectral energy distribution of \yamato\ is consistent with that of an extreme \hb+\oiii\ emitter, with equivalent width of $\sim\ew$\,\AA. Such a line emitter is consistent with an increasing trend of \oiii\ emission for high-$z$ galaxies, and can be explained by high gas temperature, a large ionization parameter, and moderate oxygen abundance.

\item By combining the non-detection of \ly\ and the high \hb+\oiii\ equivalent width inferred from \spit\ photometry, we placed an upper limit to \yamato's \ly\ escape fraction of $\simlt \fesc \%$ at $5\sigma$ confidence level. 

\item {The final interpretation of the nature of our point sources is pending. Deeper spectroscopic follow-ups or future spectroscopic observations at longer wavelength should be able to reveal their physical properties.}

\item We estimated the number density of high-$z$ point sources $\sim1\times10^{-6}$\,Mpc$^{-3}$\,mag$^{-1}$ at $M_{\rm UV}\sim-23$\,mag, and presented upper limit to their number density in the luminosity range $M_{\rm UV} \simlt -20$\,mag.

\item Additional 500-orbit of \hst\ data similar to those studied in this work would provide interesting constraints on the evolution of the quasar LF in the magnitude range $M_{\rm UV}\sim-22$\,mag. In order to detect more luminous quasars at $z\sim8$ and beyond, the large volume probed by the {Roman Space Telescope} will be necessary.

\end{enumerate}

\section*{Acknowledgements}
The data presented herein were obtained at the W.~M.~Keck Observatory, which is operated as a scientific partnership among the California Institute of Technology, the University of California and the National Aeronautics and Space Administration. The Observatory was made possible by the generous financial support of the W.~M.~Keck Foundation. The authors recognize and acknowledge the very significant cultural role and reverence that the summit of Maunakea has always had within the indigenous Hawaiian community. We are most fortunate to have the opportunity to conduct observations from this mountain. 

We thank the anonymous referee for reading the manuscript carefully and providing constructive comments in a timely manner despite this pandemic situation. The authors are grateful to Mariska Kriek and the MOSDEF team for developing and kindly letting us use their MOSFIRE pipeline, Elena Manjavacas for kindly supporting our observing run, Marco Chiaberge, Colin Norman, and Max Gronke for providing constructive comments on the manuscript, and Serena Manti for providing the data points of their LF models. Support for this work was provided by NASA through grant Nos. HST-GO-15212.002, HST-GO-15702.002, and HST-AR-15804.002-A from the Space Telescope Science Institute, which is operated by AURA, Inc., under NASA contract NAS 5-26555. MT acknowledges the support provided by the Australian Research Council Centre of Excellence for All Sky Astrophysics in 3 Dimensions (ASTRO 3D), through project No. CE170100013. 

{\it Software:} Astropy \citep{muna16}, python-fsps \citep{foreman14}.

\bibliographystyle{apj}
\bibliography{./adssample}
\end{document}